\documentclass[twocolumn]{revtex4}
\usepackage{graphicx}
\usepackage{bm}
\usepackage{graphicx}
\usepackage{amsmath}
\usepackage{mathrsfs}
\usepackage{ulem}
\usepackage{color}
\usepackage[colorlinks, linkcolor=red,anchorcolor=green,citecolor=blue]{hyperref}

\renewcommand\sout{\bgroup \color{red} \ULdepth=-.5ex \ULset}

\begin{document}
\title{ Exotic Quantum States for Charmed Baryons at Finite Temperature  }
\author{Jiaxing Zhao, Pengfei Zhuang}
\address{Physics Department, Tsinghua University and Collaborative Innovation Center of Quantum Matter, Beijing 100084, China}
\date{\today}

\begin{abstract}
The significantly screened heavy-quark potential in hot medium provides the possibility to study exotic quantum states of three-heavy-quark systems. By solving the Schr\"odinger equation for a three-charm-quark system at finite temperature, we found that, there exist Borromean states which might be realized in high energy nuclear collisions, and the binding energies of the system satisfy precisely the scaling law for Efimov states in the resonance limit.
\end{abstract}

\maketitle

Exotic states like Borromean ring~\cite{borromean} and Efimov effect~\cite{efimov1} are a kind of universal low-energy quantum phenomena for non-relativistic few-body systems. For instance, when a two-body bound state is around the scattering threshold in the unitary limit with infinite scattering length, there exist arbitrarily shallow three-body states with invariant energy spectrum under a discrete scaling transformation~\cite{efimov2,braaten}. While these Efimov states are pointed out in 1970s, it is not easy to experimentally realize them due to the difficulty of adjusting the scattering length in typical systems. The first evidence for the Efimov states was reported in an ultracold gas of caesium atoms in 2006~\cite{kraemer}, by controlling the scattering length through turning the external magnetic field $B$ to a Feshbach resonance.

The exotic states in nuclear and particle physics are widely studied in systems close to the unitary limit like few-nucleon systems, halo nuclei and weakly bound hadronic molecules, see for instance the review \cite{hammer}. Considering the necessary condition of short-range two-body potential for the exotic states, it is impossible for a three-quark system to be in exotic states in vacuum, due to the long-range two-quark interaction required by quark confinement. However, the quark interaction strength and region are significantly reduced at finite temperature and density, and there exists a phase transition from quark confinement to deconfinement at a critical temperature $T_c\simeq 160$ MeV~\cite{lattice1}. The new phase above $T_c$ is called quark-gluon plasma (QGP) and can be created in high energy nuclear collisions~\cite{qgp}. For tightly bound quarkonia made of a pair of heavy quarks, the yield suppression due to the color screening effect on the heavy quark potential has long been considered as a probe of the QGP formation in heavy ion collisions~\cite{satz,pbm,thews,rapp,zhuang}. For triply charmed baryons in QGP, the strongly screened potential between any two heavy quarks may lead to exotic baryon states around the temperature where any diquark state disappears. The temperature which is used to control the scattering length here plays the similar role as the magnetic field in cold atom gas. In this paper we study the exotic states of triply charmed baryons at finite temperature with potential model. The ground state of such baryons is $\Omega_{ccc}$. While it is not yet discovered in elementary collisions due to the small production cross section, its yield can be extremely enhanced in high energy nuclear collisions at LHC energy~\cite{he,zhao}.

To study the possible exotic states of triply charmed baryons, we first look for the unitary limit of the corresponding diquark system at finite temperature. From the lattice simulation~\cite{lattice2}, the free energy $F_{c\bar c}$ between a pair of charm quarks at finite temperature can be parameterized as~\cite{lattice3}
\begin{eqnarray}
\label{f}
F_{c\bar c}(r,T) &=& {\sigma \over \mu}\left[ {\Gamma(1/4) \over 2^{3/2}\Gamma(3/4)}-{\sqrt{\mu r} \over 2^{3/4}\Gamma(3/4)}K_{1\over4}(\mu^2r^2) \right]  \nonumber \\
&&-\alpha \left[ \mu+{e^{-\mu r}\over r} \right],
\end{eqnarray}
where $\Gamma$ is the Gamma function, $K_{1/4}$ the modified Bessel function of the second kind, and $\mu(T)$ the temperature dependent screening mass extracted from the lattice simulation. The free energy is not the potential in general case. The potential should be in between the two limits, the free energy $F_{c\bar c}$ and the internal energy $U_{c\bar c}=F_{c\bar c}+TS_{c\bar c}=F_{c\bar c}-T\partial F_{c\bar c}/\partial T$. Considering that the condition to form a three-body exotic state is a short range and strong enough force between any two particles, the formation becomes easier for $V_{c\bar c}=F_{c\bar c}$ in the sense of short range and for $V_{c\bar c}=U_{c\bar c}$ in the sense of strong strength. In the following we take the up limit $V_{c\bar c}=U_{c\bar c}$ as an example to show the formation of the exotic states. We also checked the other limit $V_{c\bar c}=F_{c\bar c}$ and found that, the exotic states still exist but the surviving temperature region is reduced. It is easy to see that $V_{c\bar c}=F_{c\bar c}=U_{c \bar c}$ in vacuum is reduced to the Cornell potential $V_{c\bar c}=-\alpha/r+\sigma r$. The three parameters in the potential model, charm quark mass $m_c=1.25$ GeV and the two coupling constants $\alpha=\pi/12$ and $\sigma=0.2$ GeV$^2$, can be fixed by fitting the masses of charmonia $J/\psi$, $\psi'$ and $\chi_c$ in vacuum. From the quark model or leading order of perturbative QCD calculation, the diquark potential is only one half of the quark-antiquark potential, $V_{cc}=V_{c\bar c}/2$. We assume such a relation at finite temperature.

\begin{figure}[htb]
{\includegraphics[width=0.45\textwidth]{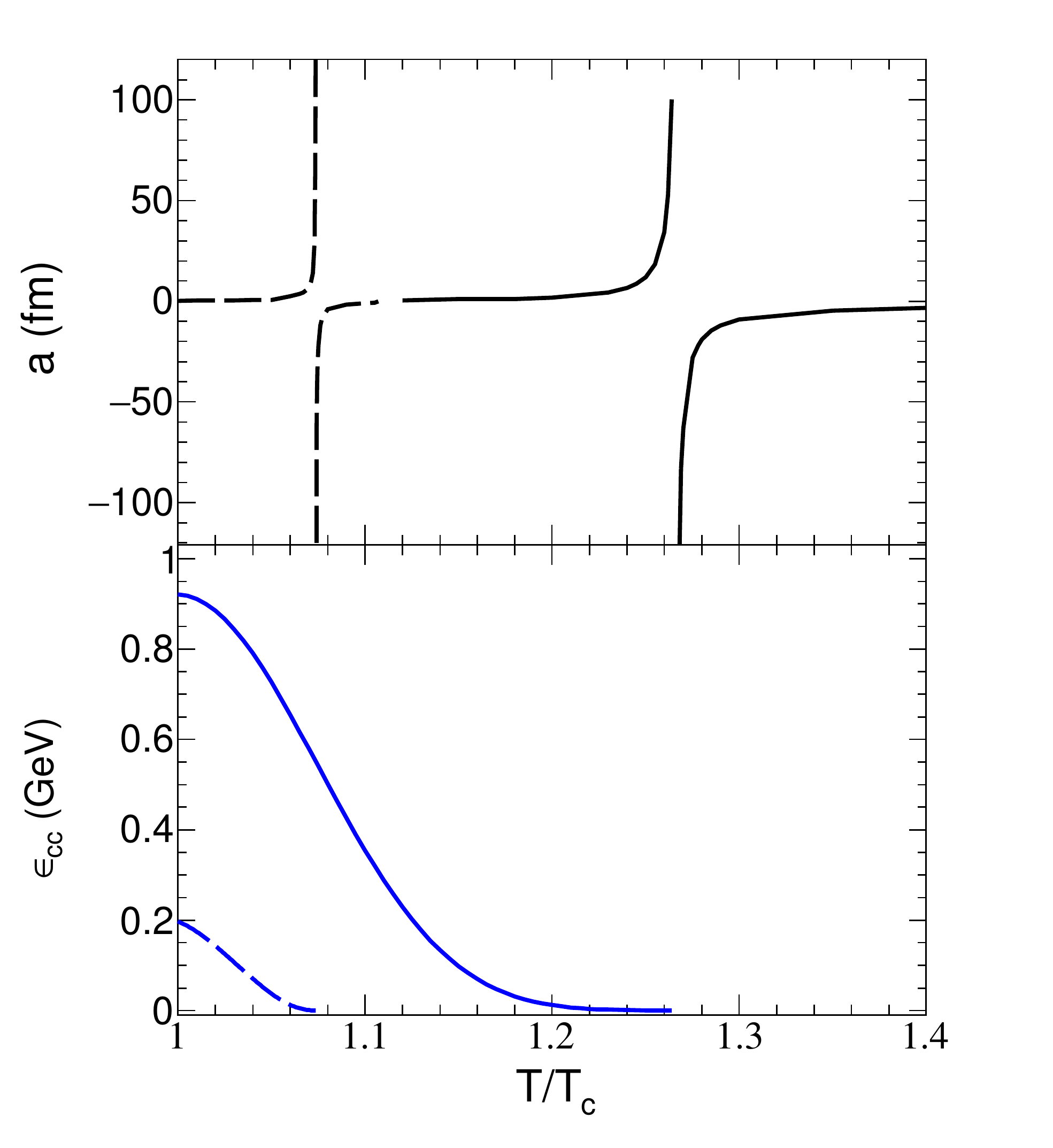}
\caption{ The scattering length $a$ and binding energy $\epsilon_{cc}$ for the ground (solid lines) and first excited (dashed lines) states of the diquark system $cc$ above the critical temperature $T_c$ of deconfinement phase transition. }
\label{fig1}}
\end{figure}

Substituting the potential $V_{cc}(r,T)$ into the radial equation for the relative motion between two charm quarks, we obtain the relative energy $E_{cc}(T)$ and in turn the binding energy $\epsilon_{cc}(T)=V_{cc}(\infty,T)-E_{cc}(T)$. It is well-known that, the low-energy S-wave scattering is described by the scattering phase shift $\delta_0(k)$, and the first term of its expansion around the zero energy $k=0$ is defined as the scattering length $a$. Around the zero-energy for S-wave, the radial wave function of the diquark system satisfies the Schr\"odinger equation
\begin{equation}
\label{radial}
\left(-{1\over m_c}{d^2\over dr^2}+V_{cc}(r,T)\right)\phi=0.
\end{equation}
Comparing the asymptotic wave function with the standard form $\phi(r,T)\approx 1-r/a(T)$ at $r\to \infty$~\cite{flambaum}, we obtain the scattering length $a(T)$. The binding energy and scattering length above $T_c$ are shown in Fig.\ref{fig1}. Different from usual nuclear systems where we can not artificially adjust the interaction to reach the resonance limit, the temperature of the hot medium controls the interaction strength and region between two heavy quarks. For the ground and first excited states, the scattering length becomes infinity at $T_d\simeq 1.27 T_c$ and $1.07 T_c$ where the corresponding binding energy approaches to zero. Below $T_d$ the attractive interaction is still strong enough to bind the two charm quarks together with positive scattering length and nonzero binding energy, while above $T_d$ the interaction becomes too weak to bind the two quarks and the scattering length becomes negative. Usually we call $T_d$ the dissociation temperature in hot medium.

We now turn to the triply charmed baryons. The wave function is governed by the Schr\"odinger equation
\begin{eqnarray}
&& \hat H \Psi({\bf r}_1,{\bf r}_2, {\bf r}_3,T) = E(T) \Psi({\bf r}_1,{\bf r}_2, {\bf r}_3,T), \nonumber\\
&& \hat H = \sum_{i=1}^3{\hat{\bf p}_i^2 \over 2m_c}+V_{ccc}({\bf r}_1,{\bf r}_2, {\bf r}_3,T),
\end{eqnarray}
where the three-quark potential is expressed as a sum of pair interactions $V_{ccc}({\bf r}_1,{\bf r}_2,{\bf r}_3,T)=\sum_{i<j} V_{cc}(|{\bf r}_i-{\bf r}_j|,T)$, and $V_{cc}$ is taken as the internal energy extracted from the lattice simulation.

It is hard to solve a three-body Schr\"odinger equation exactly. Usually one needs to take some approximation to simplify the problem. An often used method is the hyperspherical approach~\cite{few}. After transforming the individual coordinates ${\bf r}_i$ to the new coordinates ${\bf R}, {\bf r}_x$ and ${\bf r}_y$, the motion of the three-quark system is factorized into a global motion and a relative motion, $\Psi({\bf R},{\bf r}_x,{\bf r}_y)=\Theta({\bf R})\Phi({\bf r}_x,{\bf r}_y)$. Then we can introduce the hyperradius $r=\sqrt{r_x^2+r_y^2}$, hyperpolar angle $\alpha=\arctan (r_x/r_y)$ and azimuthal angles $\theta_x, \varphi_x, \theta_y$ and $\varphi_y$ to replace the relative coordinators ${\bf r}_x$ and ${\bf r}_y$. Considering the fact that the potential $V_{ccc}(r,\Omega,T)$ depends on both the hyperradius $r$ and the 5 angles $\Omega=\{\alpha,\theta_x,\phi_x,\theta_y,\phi_y\}$, the relative motion cannot be factorized into a radial part and an angular part. When the three constituents are the same, like triply charmed baryons here, one usually takes the angles averaged potential~\cite{simonov}
\begin{equation}
V_{ccc}(r,T) = {48\over \pi}\int_0^{\pi/2}V_{cc}(\sqrt 2 r\sin\alpha,T)\cos^2\alpha\sin^2\alpha d\alpha
\end{equation}
to replace $V_{ccc}(r,\Omega,T)$. Under this approximation, the factorization can be done, $\Phi(r, \Omega)=\varphi(r)Y(\Omega)$, the radial wave function $\varphi(r,T)$ and the relative energy $E_{ccc}(T)$ for the $S$-wave are controlled by the radial equation
\begin{equation}
\left[{1\over 2m_c}(-{d^2 \over dr^2}-{5\over r}{d \over dr})+V_{ccc}(r,T)\right]\varphi=E_{ccc}\varphi,
\end{equation}
and the angular part $Y(\Omega)$ is the eigenstate of the hyper angular momentum operator. By solving the radial equation at finite temperature, we obtain the binding energy of the three quark system $\epsilon_{ccc}(T) = V_{ccc}(\infty, T)-E_{ccc}(T)$.

From Fig.\ref{fig1}, the diquark bound state $cc$ can exist at temperature $T<T_d=1.27T_c$. Therefore, in this temperature region the triply charmed baryons can be in three-quark state $ccc$ or quark-diquark state $c(cc)$. For the latter, we can first solve the two-body Schr\"odinger equation for the bound state $cc$ and obtain the diquark mass $m_{cc}(T)=2m_c-\epsilon_{cc}(T)$, and then solve the two-body equation for the bound state $c(cc)$ with reduced mass $m_c m_{cc}/(m_c+m_{cc})$ and obtain the binding energy $\epsilon_{c(cc)}(T)=V_{c(cc)}(\infty,T)-E_{c(cc)}(T)$, where $E_{c(cc)}$ is the relative energy between the quark $c$ and diquark $cc$. Considering the fact that the quark-diquark system is color neutral, we have the potential $V_{c(cc)}=V_{c\bar c}$.

\begin{figure}[htb]
{\includegraphics[width=0.45\textwidth]{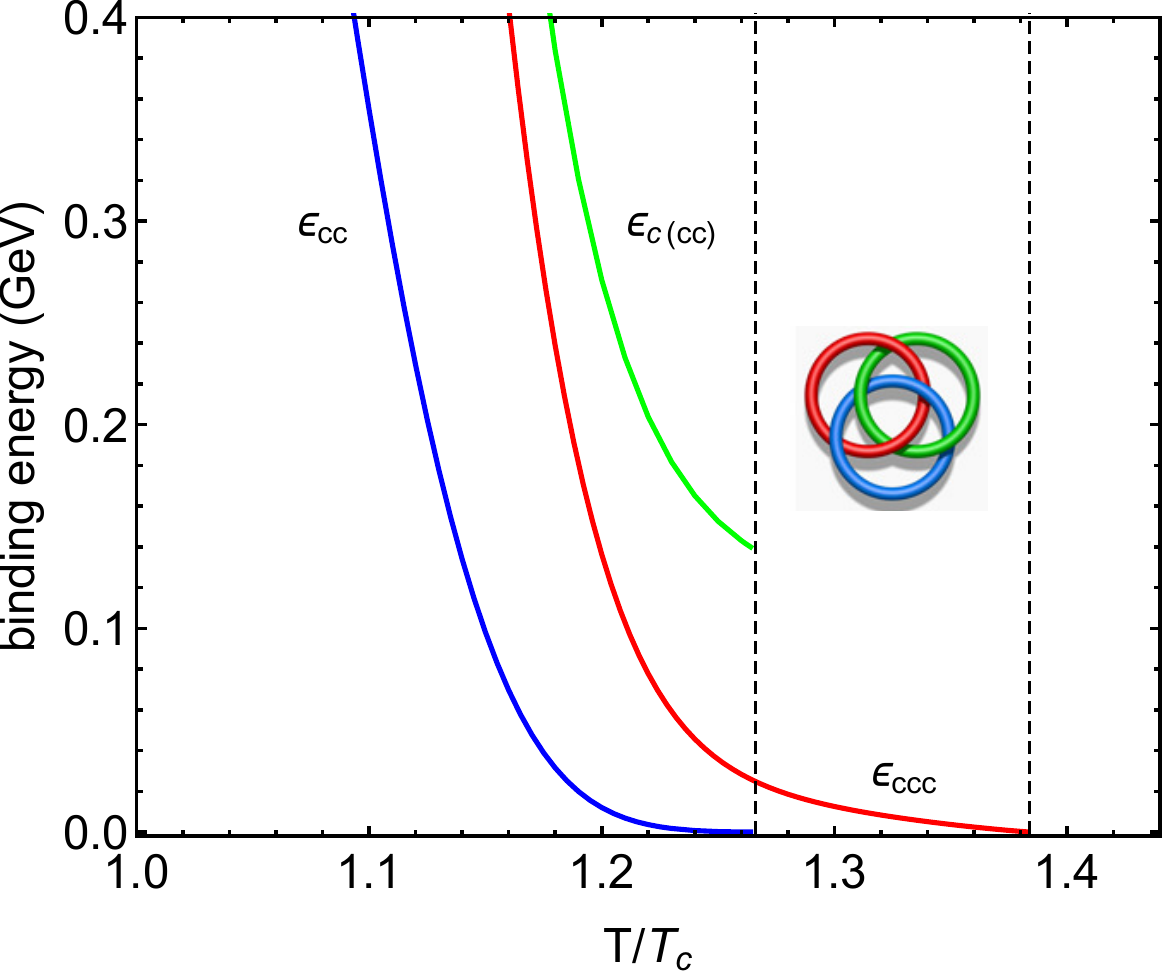}
\caption{The binding energies $\epsilon_{cc}$, $\epsilon_{ccc}$ and $\epsilon_{c(cc)}$ for the diquark, three-quark and quark-diquark states, calculated through hyperspherical method. }
\label{fig2}}
\end{figure}

The binding energies of the ground states for the diquark, quark-diquark and three-quark systems $\epsilon_{cc}$, $\epsilon_{c(cc)}$ and $\epsilon_{ccc}$ are shown in Fig.\ref{fig2} as functions of temperature. $\epsilon_{cc}(T)$ is the same as in Fig.\ref{fig1}. In the temperature region $T<T_d$ where the diquarks can survive, while both $\epsilon_{c(cc)}(T)$ and $\epsilon_{ccc}(T)$ are not zero, the quark-diquark system is a more deeply bound state than the three-quark system, $\epsilon_{c(cc)}(T) > \epsilon_{ccc}(T)$. Therefore, the ground state of triply charmed baryons in this region is the quark-diquark state $c(cc)$. In the higher temperature region $T_d < T < T_t$ where $T_t=1.38 T_c$ is the dissociation temperature of the three-quark system with binding energy $\epsilon_{ccc}\to 0$, the diquark state disappears, and triply charmed baryons can only be in the three-quark state. This shows that, three charm quarks can bind even when the interaction is too weak to bind any two charm quarks. This is the Borromean state at quark level. When the medium is above the dissociation temperature $T_t$, any triply charmed baryon state disappears, and there are only free charm quarks in the system.

The hyperspherical method used above and other versions like low-energy Faddeev equation~\cite{fedorov,faddeev} is widely used to investigate the ground state properties of three-body systems. However, to be more precise for the excited states which should be included in the study of Efimov effect, one needs to solve many coupled equations and the numerical calculation becomes rather complicated~\cite{zhao}. On the other hand, the Efimov states are defined around the scattering threshold, and we need only the solution of the three-body equation in the vicinity of the unitary limit. One of the approaches in this limit is the Separable model~\cite{yam,ernst}. It works very well in the region with large scattering length and has been checked with Van der Waals potential~\cite{pascal2}. In this model, we construct a separable potential operator to replace the real potential between two charm quarks,
\begin{equation}
\label{potential}
\hat V={\xi \over m_c} |\chi \rangle \langle \chi|.
\end{equation}
The state $|\chi \rangle$ in momentum space is expressed as
\begin{equation}
\label{chi}
\chi(q)=1-q\int_0^\infty \left(1-{r \over a}-\phi(r)\right)\sin(qr)dr
\end{equation}
and can be considered as the deviation of the real zero-energy wave function from its standard asymptotic form, where $\phi(r)$ satisfies the two-body Schr\"odinger equation (\ref{radial}) in low energy limit. The coefficient $\xi$ is chosen as
\begin{equation}
{1\over \xi} = {1\over 4\pi a}-{1\over 2\pi^2}\int_0^\infty |\chi(q)|^2 dq.
\end{equation}

The potential (\ref{potential}) has the advantage of being easily tractable because of its separability. Replacing the real potential between two heavy quarks by the separable potential in the three-quark Schr\"odinger equation in momentum space and following the simplification shown in \cite{pascal} around the unitary limit, the equation with spherical symmetry is reduced to a one-dimensional integral equation which is similar to the Skorniakov-Ter-Martirosian equation for contact potential~\cite{lee,stm1,stm2},
\begin{equation}
\label{integral}
D(P)F(P)+\int_0^\infty {q^2dq \over 2\pi^2}H(P,q)F(q)=0
\end{equation}
with
\begin{eqnarray}
\label{dh}
&&D(P) \\
=&&{1\over 4\pi a}+\int_0^\infty {dq \over 2\pi^2} \left ({ q^2|\chi(q)|^2 \over q^2-(m_cE - {3\over4}P^2)}-|\chi(q)|^2 \right), \nonumber\\
&&H(P,q) \nonumber \\
=&&\int_{-1}^1du{ \chi^*(\sqrt{q^2+{P^2\over 4}+qPu})  \chi(\sqrt{P^2+{q^2\over4}+qPu})\over P^2+q^2+qPu-m_cE},\nonumber
\end{eqnarray}
where $P$ and $E$ are respectively the relative momentum and energy of the triply charmed baryons.

By solving the integral equation (\ref{integral}) we obtain the binding energy $\epsilon_{ccc}(T)=V(\infty,T)-E(T)$ and the corresponding eigenvector $F(P)$, the latter can further reproduce the three-body wave function. The method we used to deal with the integral equation is called Nystrom method~\cite{nystrom}. With this method, the integral equation is transformed to a matrix eigen equation, and one needs to chose some approximate quadrature. In our treatment we take the Gauss-Legendre quadrature~\cite{nystrom}, and the dimension of the matrix is larger than 200.
\begin{figure}[htb]
{\includegraphics[width=0.45\textwidth]{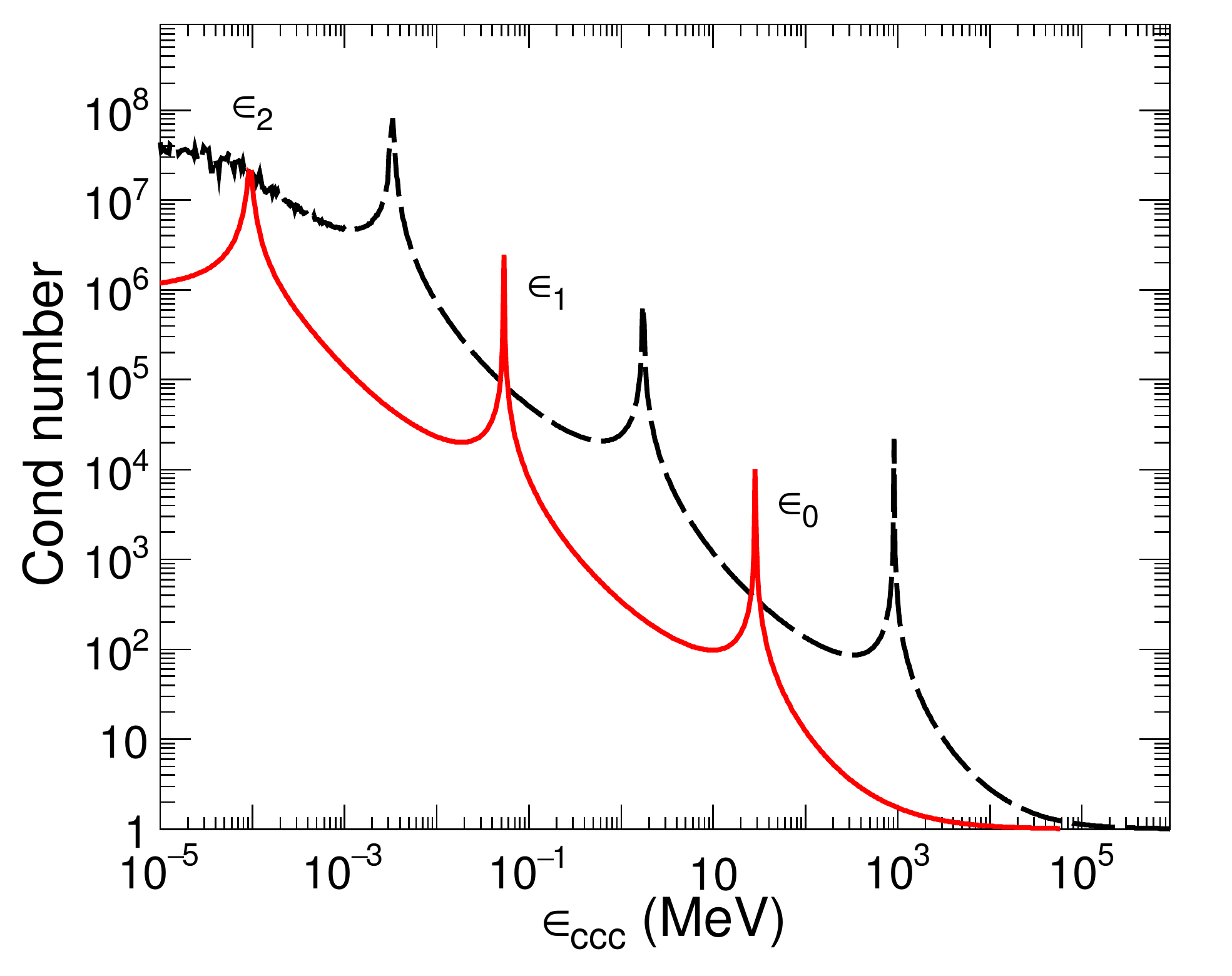}
\caption{The binding energy $\epsilon_{ccc}$ for a three-particle system with contact interaction (dashed line) and triply charmed baryons with internal energy $U(r,T_d)$ as the two-quark potential in resonance limit (solid line). }
\label{fig3}}
\end{figure}

We first take the zero-range potential (renomalizable contact potential) $V(r)=g\delta(r)$ to check the numerical method. Considering the exact solution $\phi(r)=1-r/a$ in this case, one obtains $\chi(q)=1$ from (\ref{chi}) and analytical expressions for $D$ and $H$ from (\ref{dh}). The binding energy $\epsilon$ for the three-particle system in resonance limit ($a=\infty$) is shown as dashed line in Fig.\ref{fig3}. When the cond number of the matrix becomes divergent, the corresponding $\epsilon$ is the value that satisfies the integral equation (\ref{integral}). We see clearly that, the binding energies for the first three states follow the scaling law very well, $\epsilon_n/\epsilon_{n+1}=e^{2\pi/ s_0} = 22.7^2=515$~\cite{efimov1}.

We now take the internal energy $U(r,T)$ simulated by lattice QCD as the potential between two charm quarks to calculate the low-energy wave function $\phi(r,T)$ at the dissociation temperature $T_d=1.27\ T_c$ and then solve the integral equation (\ref{integral}) for triply charmed baryon states. The obtained binding energy $\epsilon$ is shown as solid line in Fig.\ref{fig3}. For the ground state, the binding energy $\epsilon_0=27.57$ MeV is almost the same as the value shown in Fig.\ref{fig2} calculated by the hyperspherical method. This indicates that, the hyperspherical method is good enough for the ground state. However, it becomes rather difficult to precisely solve the excited states by using the hyperspherical method. In the Separable model, the binding energies are $\epsilon_1 = 0.0535$ MeV and $\epsilon_2=0.0001035$ MeV for the first and second excited states, and we have very precisely the scaling law among the ground and excited states $\epsilon_0/\epsilon_1 \approx \epsilon_1/\epsilon_2 \approx 515$. We also calculated the energy spectrum around the unitary limit where the scattering length is quite large. The scaling law is still satisfied reasonably well. These results confirm clearly the existence of Efimov effect in triply charmed baryon states at finite temperature.
\begin{figure}[htb]
{\includegraphics[width=0.45\textwidth]{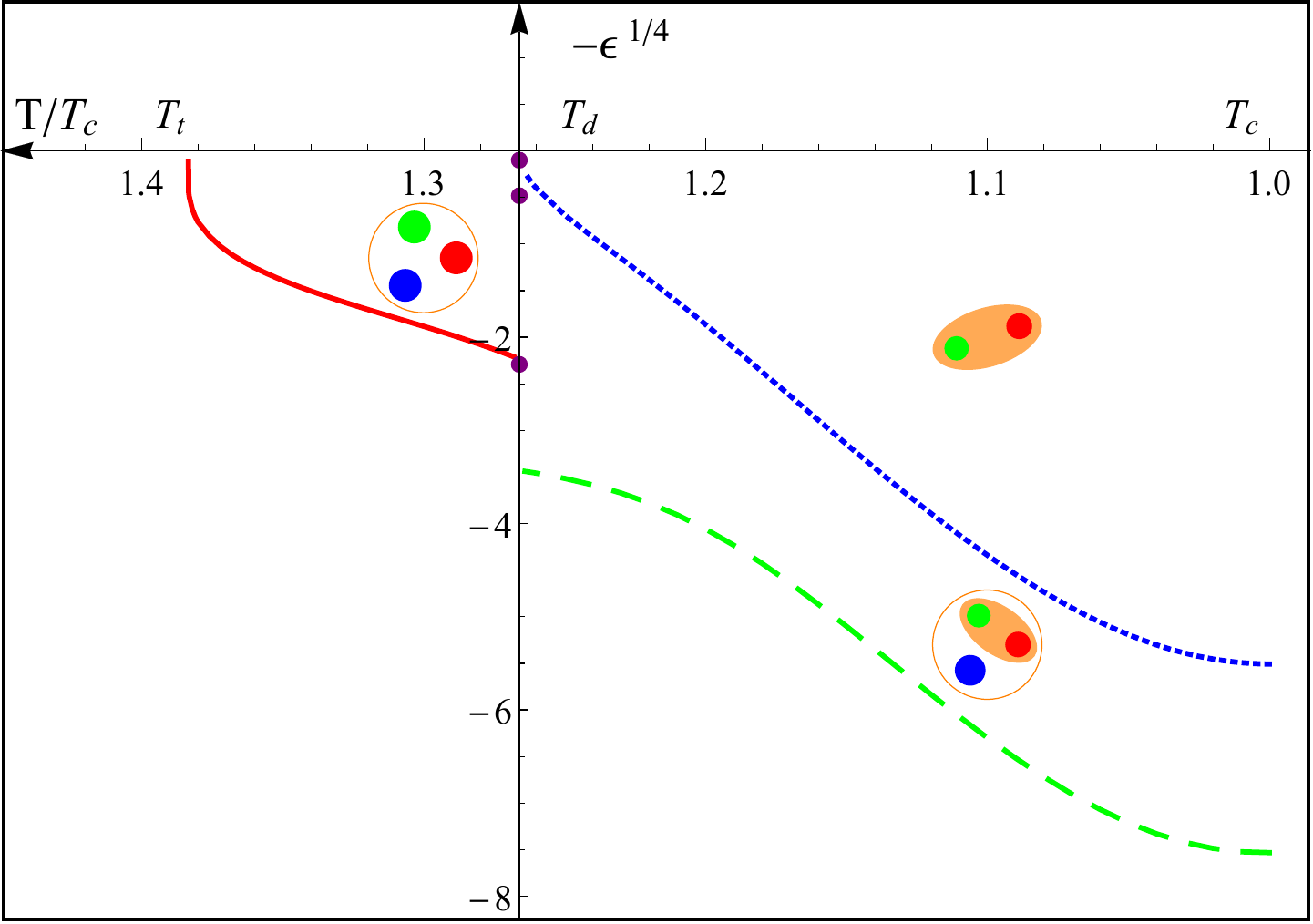}
\caption{ The scaled binding energies as functions of temperature. The dotted, dashed and solid lines correspond respectively to the ground states of diquark, quark-diquark and three-quark systems, and the three dots on the vertical axis are for the first three Efimov states in resonance limit. }
\label{fig4}}
\end{figure}

The Efimov effect is normally illustrated by showing the three-body binding energy as a function of two-body scattering length $a$~\cite{braaten}. Considering the fact that temperature controls the interaction strength and region, and using the relation between $a$ and $T$ shown in Fig.\ref{fig2}, we show in Fig.\ref{fig4} the binding energy of the ground state as a functions of $T$ for diquark and three-quark systems above $T_c$. With increasing temperature, the interaction between two heavy quarks becomes more and more weak, the diquark can only exist in the region of $T<T_d=1.27\ T_c$, but the surviving region for triply charmed baryons is extended to $T<T_t=1.38 T_c$. Above $T_t$, both the diquark and three-quark systems are dissociated in the hot medium and there are no more charmed hadrons. When the diquark survives, the ground state of triply charmed baryons prefers to be the quark-diquark state due to the much larger binding energy in comparison with the three-quark state, $\epsilon_{c(cc)} > \epsilon_{ccc}$. When diquarks are dissociated above $T_d$, the ground state can only be the three-quark state which is an exotic quantum state, namely the Borromean ring. Around the transition point from quark-diquark state to three-quark state in the unitary limit, the triply charmed baryons are in Efimov states. The binding energies for the first three Efimov states calculated through the Separable model are shown as three dots on the vertical axis in Fig.\ref{fig4}, and they satisfy precisely the scaling law $\epsilon_n/\epsilon_{n+1}=515$.

How can we realize the exotic states of triply charmed baryons in typical systems? Since the production yields of multi-charmed baryons like $\Xi_{cc}$ and $\Omega_{ccc}$ are tremendously enhanced in relativistic heavy ion collisions in comparison with elemental collisions~\cite{he,zhao}, it is most probable to discover these exotic states in nuclear collisions at RHIC and LHC energies. During the expansion of the fireball created in the early stage of heavy ion collisions, the fireball cools down. When the temperature reaches $T_t$, $\Omega_{ccc}$ can be produced via coalescence mechanism and is in the Borromean state. When the fireball reaches the resonance limit at $T_d$, the triply charmed baryons can in principle be in Efimov states. However, considering the size of the excited states, it looks hard for the Efimov states to be formed in heavy ion collisions. From the hyperspherical method the averaged radius of the ground state ($\Omega_{ccc}$) is $\langle r \rangle_0 \sim 0.7$ fm, and from the scaling law $\langle r \rangle_{n+1}/\langle r\rangle_n=\sqrt{515}$ obtained with the Seprable model, the average size of the first excited state is $\langle r\rangle_1 \sim 16$ fm which is already beyond the estimated size ($\sim 10$ fm) of the quark-gluon plasma created at RHIC and LHC.

Note that, our treatment of exotic states here is at mean field level and we did not take into account their decay by thermal quarks and gluons in hot medium. An effective way to consider the thermal decay is to add an imaginary potential to the Schr\"odinger equation. This was done for charmonium states by lattice people~\cite{burnier}. They found that, the imaginary part results in a broadening of the $J/\psi$ spectral function, but the location of the peak is not changed. This leads to a $10\%$ change in the $J/\psi$ dissociation temperature. Such effect will surely change the survival of the exotic states. It can be expected that, the surviving temperature region will be reduced when the imaginary part is included.

In summary, we investigated exotic quantum states at quark level. Since the interaction between two heavy quarks is significantly reduced from a confinement potential in vacuum to a short-range one above the critical temperature of deconfinement, it becomes possible to search for the exotic states for triply charmed baryons at finite temperature. We found that, there exists a temperature region where three charm quarks are in bound state when the attractive interaction is too weak to bind any two charm quarks. This indicates the formation of Borromean state at quark level. We also calculated the binding energies for the ground and excited states of triply charmed baryons near the resonance limit, they satisfy the scaling law among Efimov states.

\appendix {\bf Acknowledgement}: We thank Peng Zhang and Ren Zhang for helpful discussions. The work is supported by the NSFC and MOST grant Nos. 11335005, 11575093, 2013CB922000 and 2014CB845400.

\end{document}